\documentclass[aps,prl,showpacs,twocolumn,amsmath,amssymb,superscriptaddress,footinbib]{revtex4}

\usepackage[english]{babel}
\usepackage{latexsym}
\usepackage{graphics}
\usepackage{subfigure}
\usepackage{epsfig}
\usepackage{color}

\def\be{\begin{equation}}
\def\ee{\end{equation}}
\def\bea{\begin{eqnarray}}
\def\eea{\end{eqnarray}}
\def\bi{\begin{itemize}}
\def\ei{\end{itemize}}
\def\bin{\begin{enumerate}}
\def\ein{\end{enumerate}}

\def\la{\langle}
\def\ra{\rangle}

\newcommand{\vect}[1]{\mathbf{#1}}

\newcommand{\ie}{{\it i.e.}~}

\newcommand{\compl}[1]{{\overline{#1}}}

\newcommand{\spin}{\mathbf{\sigma}}
\newcommand{\h}{\mathbf{h}}

\begin{document}

\title{Disorder-Induced Order in Two-Component Bose-Einstein Condensates}

\author{A.~Niederberger} \affiliation{ICFO-Institut de Ci\`encies Fot\`oniques, Parc Mediterrani de la Tecnologia, E-08860 Castelldefels (Barcelona), Spain}
\author{T.~Schulte} \affiliation{ICFO-Institut de Ci\`encies Fot\`oniques, Parc Mediterrani de la Tecnologia, E-08860 Castelldefels (Barcelona), Spain}\affiliation{Institut f\"ur Quantenoptik, Leibniz Universit\"at Hannover, D-30167 Hannover, Germany}
\author{J.~Wehr} \affiliation{ICFO-Institut de Ci\`encies Fot\`oniques, Parc Mediterrani de la Tecnologia, E-08860 Castelldefels (Barcelona), Spain} \affiliation{Department of Mathematics, The University of Arizona, Tucson, AZ 85721-0089, USA}
\author{M.~Lewenstein} \affiliation{ICFO-Institut de Ci\`encies Fot\`oniques, Parc Mediterrani de la Tecnologia, E-08860 Castelldefels (Barcelona), Spain} \affiliation{ICREA - Instituci\`o Catalana de Ricerca i Estudis Avan{\c c}ats, E-08010 Barcelona, Spain}
\author{L.~Sanchez-Palencia} \affiliation{Laboratoire Charles Fabry de l'Institut d'Optique, CNRS and Univ. Paris-Sud, Campus Polytechnique, RD 128, F-91127 Palaiseau cedex, France}
\author{K.~Sacha} \affiliation{ICFO-Institut de Ci\`encies Fot\`oniques, Parc Mediterrani de la Tecnologia, E-08860 Castelldefels (Barcelona), Spain} \affiliation{Marian Smoluchowski Institute of Physics and Mark Kac Complex Systems Research Centre, Jagiellonian University, Reymonta 4, PL-30059 Krak\'ow, Poland}

\date{\today}

\begin{abstract}
We propose and analyze a general mechanism of {\it disorder-induced order} in two-component Bose-Einstein condensates, analogous to corresponding effects established for $XY$ spin models. We show that a random Raman coupling induces a relative phase of $\pi/2$ between the two BECs and that the effect is robust. We demonstrate it in 1D, 2D and 3D at $T=0$ and present evidence that it persists at small $T>0$. Applications to phase control in ultracold spinor condensates are discussed.
\end{abstract}

\pacs{05.30.Jp, 03.75.Hh, 03.75.Mn, 64.60.Cn}

\maketitle

Degenerate quantum gases offer unprecedented control tools
\cite{Advan}, opening fascinating possibilities, e.g.
investigations of quantum disordered systems \cite{ahufinger2005}.
Current experimental \cite{inguscio,clement,schulte} and
theoretical \cite{schulte,lifshits,lsp2007,pavloff} works are
mainly devoted to studies of the interplay between disorder and
nonlinearity in Bose-Einstein condensates (BEC) in a quest for
traces of Anderson localization. In the regime of strong
correlations, evidence for the Bose glass phase has been reported
\cite{fallani}, and even more exotic quantum phases have been
proposed \cite{ahufinger2005,graham}.

Weak disorder can have strong effects also in classical systems.
For instance, a general mechanism of {\it random-field-induced
order} (RFIO) has been proposed recently \cite{our,ourbis}. It is
responsible for ordering in graphene quantum Hall ferromagnets
\cite{lee}, $^3$He-A in aerogel and amorphous ferromagnets
\cite{volovik}, as well as for inducing superfluidity in hardcore
bosonic systems \cite{JWLee}. This effect is best understood in
classical ferromagnetic $XY$ models in the presence of uniaxial
random magnetic fields. For the 2D square lattice, the Hamiltonian
reads
\begin{equation}
H = - \sum_{|i-j| = 1}\spin_i \cdot \spin_j
    -  \sum_i \h_i \cdot \spin_i,
\label{xy}
\end{equation}
where the spins are unit 2D vectors in the $XY$ plane:
$\spin_i=(\cos\theta_i,\sin\theta_i)$ at site $i \in {\bf Z}^2$.
When $\h \equiv 0$ the system does not magnetize as a consequence
of the Mermin-Wagner-Hohenberg (MWH) theorem \cite{mwh}. In
contrast, a weak uniaxial random field $\h_i$ breaks the
continuous $U(1)$ symmetry. Then, the MWH theorem does not apply
and the system spontaneously magnetizes with a non-zero component
of the magnetization perpendicular to the random field. This has
been proven at zero temperature and strong arguments have been
given that the effect persists at small temperatures \cite{our}.
Hamiltonian~(\ref{xy}) can be realized with ultracold atoms in
optical lattices, but the effect is rather weak \cite{our}.

In this Letter, we propose an analogue of the RFIO effect using
two BECs trapped in harmonic potentials and coupled via a
real-valued random Raman field. We show that the mean field
Hamiltonian of the two-component BEC is analogous to the $XY$ spin
Hamiltonian~(\ref{xy}), with the Raman coupling playing the same
role as the magnetic field in Eq.~(\ref{xy}), and the relative
phase between the BECs corresponding to the spin angle $\theta_i$.
Then, the RFIO effect shows up in the form of a relative phase
between the BECs fixed at a value of $\pm \pi/2$. The finite-size
two-component BEC system is continuous and formally equivalent to
the discrete spin system~(\ref{xy}) on an infinite lattice. We
find that even in low dimensions, the RFIO effect is much more
pronounced and robust in coupled trapped BECs than it is in
uniform lattice spin models. Note that trapped (finite size) BECs
at sufficiently small $T$ show true long range order also in 1D
and 2D as phase fluctuations take place on a scale larger than the
size of the systems \cite{Gora}. We demonstrate the effect in 1D,
2D and 3D at $T=0$ and present strong evidence that it persists
for small $T>0$.

Interestingly, the RFIO effect is quite general. For instance,
consider the two-spin lattice Hamiltonian:
\begin{equation}
H = -\sum_{|i-j|=1} (\mathbf{\sigma}_i\cdot\mathbf{\sigma}_j +
\mathbf{\tau}_i \cdot \mathbf{\tau}_j)
    - \sum_i \Omega_i \mathbf{\sigma}_i \cdot \mathbf{\tau}_i,
\label{xybis}
\end{equation}
where $\Omega_i$ are independent real-valued random couplings with
(identical) symmetric distributions. In this system, it can be
proven rigorously that there is no first order phase transition
with the order parameter $\mathbf{\sigma}_i \cdot \mathbf{\tau}_i$
in dimensions $d \leq 4$ \cite{aizenman}. More precisely, in every
infinite-dimensional Gibbs state (phase), the disorder average of
the thermal mean $\la\mathbf{\sigma}_i \cdot \mathbf{\tau}_i\ra$
takes the same value. By symmetry, this value has to be zero,
implying that the average cosine of the angle between
$\mathbf{\sigma}_i$ and $\mathbf{\tau}_i$ is zero. At $T=0$, these
results also apply \cite{aizenman} and are consistent, by analogy,
with the relative phase $\pi/2$ of two randomly coupled BECs,
discussed below.

We consider a trapped two-component Bose gas with repulsive
interactions and assume that the two components consist of the same
atomic species in two different internal states, coupled via a
position-dependent (random, quasi-random, or just oscillating)
real-valued Raman field $\Omega (\vect{r})$ of mean zero ($\int
\Omega \textrm{d}\vect{r}=0$). The typical amplitude and spatial
variation scale of $\Omega (\vect{r})$ are denoted by
$\Omega_\textrm{R}$ and $\lambda_\textrm{R}$. At sufficiently
small $T$, the trapped gases form BECs which can be represented by
the classical fields $\psi_{1,2} (\vect{r})$ in the mean-field
approximation. The energy functional of the system then reads
\bea
E & = &
\int \textrm{d}\vect{r} \left[
\phantom{i}
({\hbar^2}/{2m})
|\nabla \psi_1|^2 + V (\vect{r})|\psi_1|^2 +
({g_1}/{2})|\psi_1|^4
\right. \nonumber \\
 & &
\phantom{\int \textrm{d}\vect{r}}
\left.
+
({\hbar^2}/{2m})
|\nabla \psi_2|^2 + V(\vect{r})|\psi_2|^2 +
({g_2}/{2}) |\psi_2|^4
\right. \nonumber \\
 & &
\left.
+ g_{12}|\psi_1|^2|\psi_2|^2
+
({\hbar\Omega(\vect{r})}/{2})
\left(\psi_1^*\psi_2+\psi_2^*\psi_1\right)
\right],
\label{enfun}
\eea
where $V(\vect{r})$ is the confining potential, and
$g_i=4\pi\hbar^2a_i/m$ and $g_{12}=4\pi\hbar^2a_{12}/m$ are the
intra- and inter-state coupling constants, with $a_i$ and $a_{12}$
the scattering lengths and $m$ the atomic mass. The last term in
Eq.~(\ref{enfun}) represents the Raman coupling which can change
the internal state of the atoms.

The ground state of the coupled two-component BEC system is
obtained by minimizing $E$ as a function of the fields $\psi_1$
and $\psi_2$ under the constraint of a fixed total number of atoms
$N=\int \textrm{d}\vect{r} (|\psi_1|^2+|\psi_2|^2)$. This leads to
a set of two coupled Gross-Pitaevskii equations (GPE):
\bea
\mu \psi_i & = & [
-\hbar^2\nabla^2/2m + V + g_i |\psi_i|^2 + g_{12} |\psi_\compl{i}|^2
] \psi_i \nonumber \\
&& + (\hbar \Omega/2)\psi_\compl{i},
\label{gpe}
\eea
with $\mu$ the chemical potential and $\compl{i}=2 (1)$ for $i=1 (2)$.

At equilibrium, for $\Omega_\textrm{R} = 0$ and $g_{1},g_{2} >
g_{12}$, the BECs are miscible \cite{ho1996}. Their phases
$\theta_{i}$ are uniform, arbitrary and independent.
Now, a weak Raman coupling ($\hbar|\Omega_\textrm{R}| \ll \mu$)
does not noticeably affect the densities. However, arbitrarily
small $\Omega (\vect{r})$ breaks the continuous $U(1)$ symmetry
with respect to the relative phase of the BECs and, following the
results of Refs.~\cite{our,ourbis,aizenman}, the relative phase
can be expected to be fixed. To make this clearer, we neglect the
changes of the densities when the weak Raman coupling is turned on
and analyze the phases. For simplicity we suppose $g_1=g_2$ and
$\rho(\vect{r})=\rho_1(\vect{r})=\rho_2(\vect{r})$. The
substitution $\psi_i=\textrm{e}^{i\theta_i (\vect{r})}
\sqrt{\rho(\vect{r})}$ in the energy functional~(\ref{enfun})
leads to $E=E_0+\Delta E$ where $E_0$ is the energy for
$\Omega_\textrm{R} = 0$ and
\bea
\Delta E & = &
\int \textrm{d}\vect{r}
\rho(\vect{r})\left[\frac{\hbar^2}{4m} (\nabla\theta)^2
+ \hbar\Omega (\vect{r}) \cos\theta
\right]
\nonumber \\
& + & \int \textrm{d}\vect{r}
\rho(\vect{r}) \frac{\hbar^2}{4m} (\nabla\Theta)^2,
\label{denfun}
\eea
where $\Theta=\theta_1 + \theta_2$ and $\theta = \theta_1 -
\theta_2$. Minimizing $\Delta E$ implies $\Theta=\rm const$, hence
the second line in Eq.~(\ref{denfun}) vanishes and the only
remaining dynamical variable in the model is the relative phase
$\theta$ between the BECs. Note that if $\rho_1\ne\rho_2$ the
variables $\Theta$ and $\theta$ are coupled and one cannot
consider them independent (the $\rho_1\ne\rho_2$ case is analyzed
in the sequel). Equation~(\ref{denfun}) is equivalent to the
classical field description of the spin model~(\ref{xy}) in the
continuous limit, where the relative phase $\theta (\vect{r})$
represents the spin angle and the Raman coupling $\Omega
(\vect{r})$ plays the role of the magnetic field. Thus, we expect
RFIO \cite{our} to show up in the form $\cos \theta \simeq 0$ for
weak random $\Omega (\vect{r})$.

Let us examine Eq.~(\ref{denfun}) in more detail. It represents a
competition between the kinetic term which is minimal for uniform
$\theta$ and the potential term which is minimal when the sign of
$\cos\theta$ is opposite to that of $\Omega (\vect{r})$. For
$\hbar\Omega_\textrm{R} \gg \hbar^2/2m\lambda_\textrm{R}^2$, the
potential term dominates and $\theta$ will vary strongly on a
length scale of the order of $\lambda_\textrm{R}$. In contrast, if
$\hbar\Omega_\textrm{R} \ll \hbar^2/2m\lambda_\textrm{R}^2$ the
kinetic term is important and forbids large modulations of
$\theta$ on scales of $\lambda_\textrm{R}$. The Euler-Lagrange
equation of the functional~(\ref{denfun}) is
\be
\nabla\left[\rho(\vect{r}) \nabla\theta\right] +
\frac{2m}{\hbar}
\rho(\vect{r})
\Omega(\vect{r})\sin\theta = 0.
\label{le}
\ee
For the homogeneous case ($\rho=\rm const$) and for slowly varying
densities (neglecting the term $\nabla \rho$), assuming small
variations of the relative phase, $\theta (\vect{r}) = \theta_0 +
\delta\theta (\vect{r})$ with $|\delta\theta| \ll \pi$, the
solution of Eq.~(\ref{le}) reads
\be
\delta \hat{\theta} (\vect{k}) \simeq
({2m}/{\hbar})
({\hat{\Omega} (\vect{k}) / |\vect{k}|^2})
\sin \theta_0
\label{sol}
\ee
in Fourier space. Inserting Eq.~(\ref{sol}) into Eq.~(\ref{denfun}), we find
\be
\Delta E \simeq - m\rho
\int \textrm{d}\vect{k}\ ({|\hat{\Omega} (\vect{k})|^2}/{|\vect{k}|^2})
\sin^2 \theta_0.
\label{energy}
\ee
The energy is thus minimal for $\theta_0 = \pm \pi/2$, \ie $\cos
\theta_0 = 0$. This indicates RFIO in the two-component BEC system
owing to the breaking of the continuous $U(1)$ symmetry of the
coupled GPEs. For a random Raman coupling, even if the resulting
fluctuations of $\theta$ are not small, the average phase is
locked at $\theta_0=\pm\pi/2$. Note that if $\theta (\vect{r})$ is
a solution of Eq.~(\ref{le}), so is $-\theta (\vect{r})$. This
follows from the fact that for any solution $(\psi_1,\psi_2)$ of
the GPEs ~(\ref{gpe}), $(\psi_1^*,\psi_2^*)$ is also a solution
with the same chemical potential. The sign of $\theta_0$ thus
depends on the realization of the BECs and is determined by
spontaneous breaking of the $\theta \leftrightarrow -\theta$
symmetry.

\begin{figure}
\centering
\vskip-0.3cm
\includegraphics*[width=8.6cm]{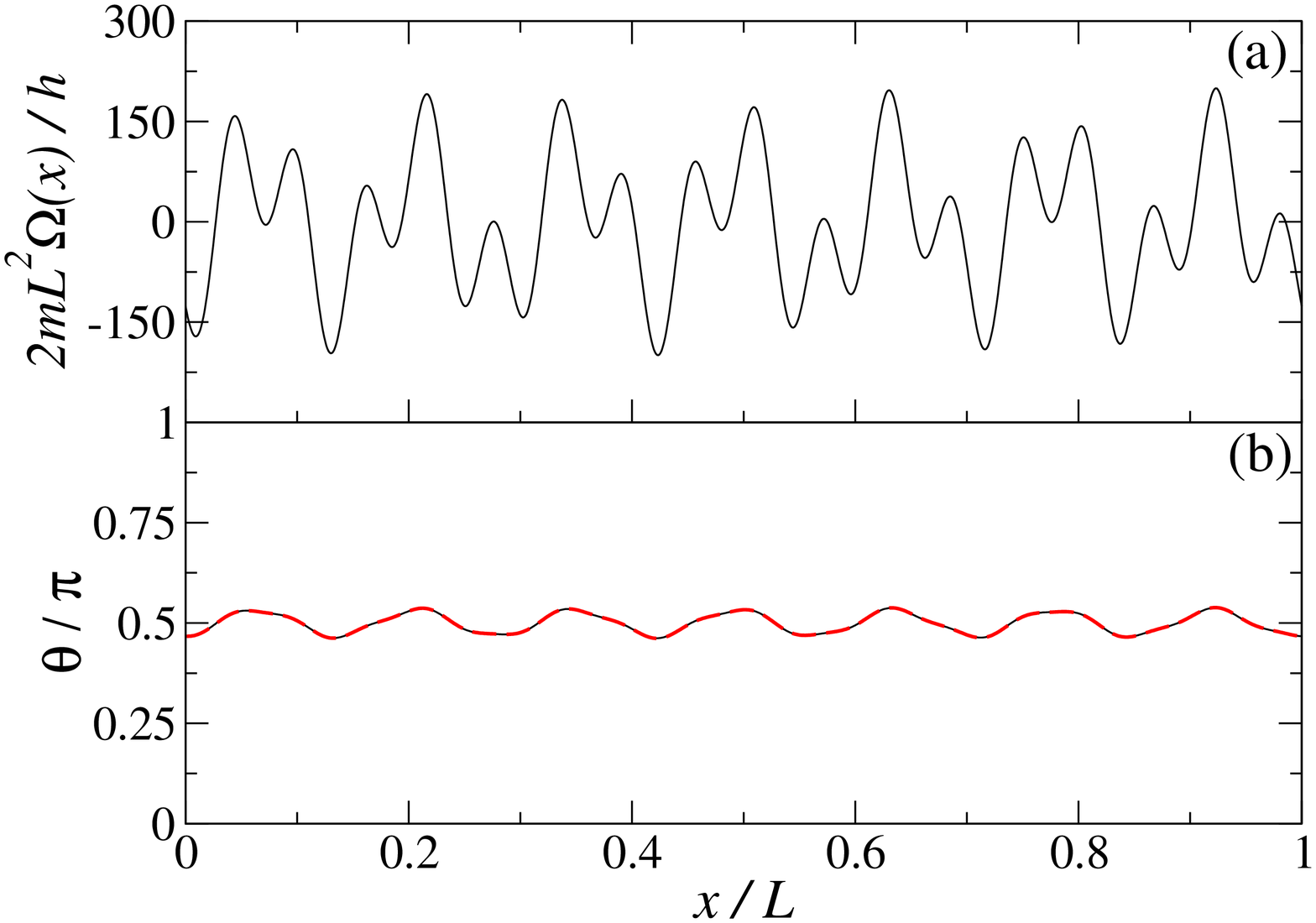}
\vskip-0.4cm \caption{RFIO effect in a 1D two-component BEC
trapped in a box of length $L$ and in a quasi-random Raman field.
Panel (a): Raman coupling function $\Omega(x) = - 100
({\hbar}/{2mL^2}) [\sin(x/\lambda_\textrm{R} + 0.31) +
\sin(x/(2.44\lambda_\textrm{R}) + 1.88)]$ with
$\lambda_\textrm{R}=0.00939 L$. Panel (b): Relative phase $\theta
(x) = \theta_1(x)-\theta_2(x)$ obtained by solving Eq.~(\ref{le})
numerically (solid black line) and comparison with Eq.~(\ref{sol})
(dashed red line --- nearly identical to the solid black line).}
\label{one}
\end{figure}

Let us turn to numerics starting with $g_1=g_2$. For homogeneous
($\rho=\rm const$) gases, we solve Eq.~(\ref{le}).
Figure~\ref{one} shows an example for a 1D two-component BEC,
where $\Omega(x)$ is a quasi-random function chosen as a sum of
two sine functions with incommensurate spatial periods. The {\it
dynamical} system (\ref{le}) is not integrable. It turns out that
the solution we are interested in corresponds to a hyperbolic
periodic {\it orbit} surrounded by a considerable chaotic sea.
Figure~\ref{one} confirms that $\theta (x)$ oscillates around
$\theta_0 \simeq \pm \pi/2$. The oscillations of $\theta(x)$ are
weak and follow the prediction~(\ref{sol}), which in 1D, after
inverse Fourier transform, corresponds to the double integral of
$\Omega(x)$.

\begin{figure}
\centering
\vskip-0.4cm
\includegraphics*[width=8.6cm]{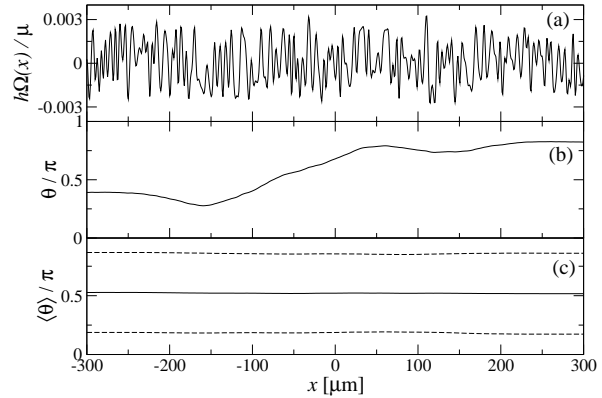}
\vskip-0.3cm \caption{RFIO effect in very elongated (effectively
1D) trapped BECs. The data corresponds to $^{87}$Rb atoms in two
different internal states in an anisotropic harmonic trap with
frequencies $\omega_x=2\pi\times 10$~Hz and
$\omega_\perp=2\pi\times 1.8$~kHz. The total number of atoms is
$N=10^4$ and the scattering lengths are $a_1=5.77$~nm,
$a_2=6.13$~nm and $a_{12}=5.53$~nm. Panel (a): Single realization
of the random Raman coupling
$\hbar\Omega/\mu$ for $\lambda_\textrm{R}=10^{-2}L_\textrm{TF}$
and $\hbar\Omega_\textrm{R} \simeq 3\times 10^{-3}\mu$. Panel (b):
Relative phase $\theta$ corresponding to $\Omega (x)$ shown in
panel (a). Panel (c): $\theta$ averaged over many realizations of
$\Omega (x)$ (solid line) and the averaged value $\pm$ standard
deviation (dashed lines). In panel (c) the solutions with
$\int\theta dx>0$ only are collected (the other class of solutions
with $\theta\rightarrow-\theta$ is not included).} \label{two}
\end{figure}

For trapped gases and for $g_1\ne g_2$ we directly solve the
coupled GPEs~(\ref{gpe}). Figure~\ref{two} shows the results for a
1D two-component BEC in the Thomas-Fermi regime confined in a
harmonic trap with a random $\Omega (x)$. A typical realization is
shown in Fig.~\ref{two}a. For each realization of $\Omega(x)$, the
resulting relative phase $\theta$ can change significantly but
only on a scale much larger than $\lambda_{\rm R}$ because
$\hbar\Omega_\textrm{R} \ll \hbar^2/2m\lambda_\textrm{R}^2$, as
shown in Fig.~\ref{two}b. However, averaging over many
realizations of the random Raman coupling and keeping only those
with $\int \theta (x) \textrm{d}x > 0$ (resp. $<0$), we obtain
$\la\theta(x)\ra \approx\pi/2$ (resp. $-\pi/2$),
with the standard deviation about $0.3\pi$ as shown in Fig.~\ref{two}c.

The dynamical stability of the solutions of the GPEs~(\ref{gpe}) found in the 1D trapped geometry can be tested by means of the Bogoliubov-de Gennes (BdG) theory which allows also to estimate the quantum depletion of the BECs \cite{StringariPitaevskii}. The BdG analysis shows that the solutions of the GPEs~(\ref{gpe}) are indeed stable and that the BdG spectrum is not significantly affected by the Raman coupling. It implies that turning on the Raman field does not change the thermodynamical properties of the system, and the RFIO effect should persist for sufficiently low $T>0$. Note that the GPEs~(\ref{gpe}) possess also a solution with both components real. However, this solution is dynamically unstable. In fact, there is a BdG mode associated with an imaginary eigenvalue and the corresponding BECs phases (under a small perturbation) will evolve exponentially in time. In addition, the BdG analysis shows that the quantum depletion is about $1\%$ and can therefore be neglected.

\begin{figure}
\centering
\vskip-0.5cm
\includegraphics*[width=8.cm]{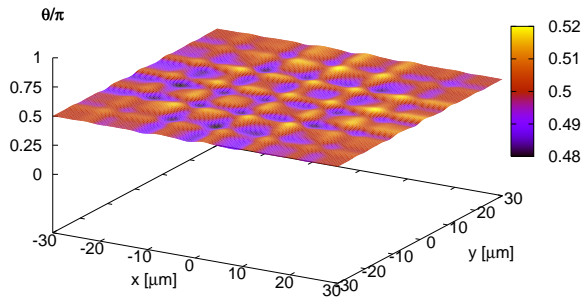}
\vskip-0.4cm \caption{RFIO effect in a 3D two-component BEC
trapped in a spherically symmetric harmonic trap with frequency
$\omega = 2 \pi \times 30 \text{Hz}$. The total number of atoms is
$N=10^5$, the scattering lengths are as in Fig.~\ref{two} and we
use a quasi-random Raman coupling $\Omega (x,y,z) \propto \sum_{u
\in \{x,y,z\}} [\sin(u/\lambda_\textrm{R}) +
\sin(u/(1.71\lambda_\textrm{R}))]$ with
$\lambda_\textrm{R}=4.68\mu\textrm{m}/2\pi$ and $\hbar
\Omega_\textrm{R} \simeq 5 \times 10^{-3}\mu$. Shown is the
relative phase $\theta$ in the plane $z=0 \mu$m in units of
$\pi$.} \label{three}
\end{figure}

Calculations in 2D and 3D --- whose detailed results will be
published soon --- show essentially the same Disorder-Induced
Ordering effect in all dimensions. For example, Fig.~\ref{three}
shows the result for two coupled 3D BECs in a spherically
symmetric harmonic trap. Here, the Raman coupling is a sum of
quasi-random functions similar to that used for Fig.~\ref{one} in
each spatial direction and with $\hbar \Omega_\textrm{R} \simeq
10^{-2}\mu$. The density modulations are found to be negligible.
However, even for this low value of the Raman coupling,
Fig.~\ref{three} shows that the relative phase is fixed around
$\theta_0=\pi/2$ with small fluctuations. Other calculations
confirm that the sign of $\theta_0$ is random but with
$|\theta_0|=\pi/2$ for all realizations of $\Omega (\vect{r})$ and
that the weaker the Raman coupling, the smaller the modulations of
$\theta (\vect{r})$ around $\theta_0$. This shows once again the
enormous robustness of RFIO in two-component BECs.

In summary, we have shown that RFIO occurs in a system of two BECs
coupled via a real-valued random Raman field. It has been
demonstrated in 1D, 2D and 3D for homogeneous or trapped BECs. The
signature of RFIO is a fixed relative phase between the BECs
around $\theta_0 = \pm \pi/2$. For quasi-random Raman coupling,
the fluctuations can be very small ($0.05\pi$ for the parameters
used in Fig.~\ref{one}). For completely random Raman coupling the
fluctuations can be larger (about $0.3\pi$ for the parameters used
in Fig.~\ref{two}). Interestingly, the two-component BEC system is
continuous and RFIO is stronger and more robust than in lattice
spin Hamiltonians of realistic sizes \cite{our}. RFIO can thus be
obtained in current experiments with two-component BECs
\cite{matthews1998,hall1998} and observed using matterwave
interferometry techniques \cite{hall1998}.

Apart from its fundamental importance, RFIO can have applications
for engineering and manipulations of quantum states by providing a
simple and robust method to control phases in ultracold gases. We
find particularly interesting applications of phase control in
spinor BECs and, more generally, in ultracold spinor gases
\cite{Advan}. For example, in a ferromagnetic spinor BEC with
$F=1$ as in $^{87}$Rb, the wavefunction is $\xi \propto
(e^{-i\phi}\cos^2(\theta/2), \sqrt{2}\sin(\theta/2)\cos(\theta/2),
e^{+i\phi}\cos^2(\theta/2))$, the components correspond to
$m_F=1,0,-1$ and the direction of magnetization is $\vec
n=(\sin\theta\cos\phi, \sin\theta\sin\phi, \cos\theta)$. Applying
two real-valued random Raman couplings between $m_F=0$ and
$m_F=\pm 1$, fixes $\phi= 0$ or $\pi$, \ie the magnetization will
be in the $XZ$ plane. By applying two random real-valued Raman
couplings between $m_F=0$ and $m_F= 1$ and between $m_F=-1$ and
$m_F= 1$, we force the magnetization to be along $\pm Z$. Similar
effects occur in antiferromagnetic spinor BECs with $F=1$, as
$^{14}$Na. Using Raman transitions with arbitrary phases,
employing more couplings, and higher spins $F$ offers a variety of
control tools in ultracold spinor gases.

We acknowledge support of EU IP Program `SCALA', ESF PESC Program
`QUDEDIS', Spanish MEC grants (FIS 2005-04627, Conslider Ingenio
2010 `QOIT'), and French DGA, IFRAF, MENRT and ANR.
J.W. was partially supported by NSF grant DMS 0623941. K.S.
acknowledges Polish Government scientific funds (2005-2008) as a
research project and M.~Curie ToK project COCOS
(MTKD-CT-2004-517186).


\end{document}